\documentstyle[12pt]{article}

   \font\tenmsb=msbm10 scaled\magstep 1
   \font\sevenmsb=msbm7 scaled \magstep 1
   \font\faivemsb=msbm5 scaled \magstep 1
\newfam\msbfam
      \textfont\msbfam=\tenmsb
      \scriptfont\msbfam=\sevenmsb
      \scriptscriptfont\msbfam=\faivemsb
\def\Bbb#1{{\fam\msbfam #1}}
\font\tengothic=eufm10 scaled\magstep 1
\font\sevengothic=eufm7 scaled\magstep 1
\newfam\gothicfam
      \textfont\gothicfam=\tengothic
      \scriptfont\gothicfam=\sevengothic

\begin{document}

\begin{center}

{\Large{\bf Principle of Pattern Selection for Nonequilibrium 
Phenomena} \\ [5mm] 

V.I. Yukalov} \\ [3mm]

{\it Bogolubov Laboratory of Theoretical Physics \\
Joint Institute for Nuclear Research, Dubna 141980, Russia\\
and \\
International Institute of Theoretical and Applied Physics \\
Iowa State University, Ames, Iowa 50011, USA}

\end{center}

\vskip 2cm

\begin{abstract}

A general principle is advanced allowing the classification of
nonunique solutions to nonlinear evolution equations, corresponding to 
different spatio-temporal patterns. This is done by defining the probability 
distribution of patterns, which characterizes multiple solutions as more or 
less probable with respect to each other. The most probable pattern is 
naturally defined by the maximum of the pattern distribution. This maximum 
is shown to be equivalent to the {\it minimum of local contraction}. The
formulated principle plays for nonequilibrium dynamical systems the same 
ordering role as the condition of minimal free energy for equilibrium 
statistical systems. The generality of the principle is illustrated by 
several examples of different nature.

\end{abstract} 

\vskip 2cm


\vskip 1cm


\newpage

\section{Introduction}

The problem of pattern selection is an old problem that has not
yet found an appropriate solution. This problem appears when a
system of nonlinear evolution equations possesses, under the same
initial and boundary conditions, several solutions describing
different spatio-temporal structures. Suppose, for convenience,
that these solutions can be parametrized by a multiparameter
$\beta$ from a manifold ${\cal B}$. It often happens that all
solutions labelled by different $\beta\in{\cal B}$ are stable,
then manifold ${\cal B}$ is called the {\it stability balloon} [1].
Since all solutions from the ensemble attached to the stability
balloon are stable, there is no general way of distinguishing
between such solutions and therefore between the related patterns.
But the necessity of distinguishing them arises because the real
life, that is assumed to be described by the corresponding
evolution equations, does distinguish between different patterns:
some of the latter appear in a given experimental protocol more
often than other. Experiments demonstrate that Nature does prefer
some patterns as more probable.

A similar problem exists in equilibrium statistical mechanics
where it often happens that a nonlinear equilibrium system
possesses several solutions for an observable quantity, say for an
order parameter. The way of treating such a nonuniqueness of
solutions for equilibrium systems is well known, being given by
the condition of the minimal free energy: More stable is that
solution and, respectively, that thermodynamic state which
corresponds to the lower free energy, the lowest free energy
defining an absolutely stable state.
But for nonequilibrium systems, there is no such a general ordering
principle permitting one to distinguish between more and less
probable solutions. This problem  of pattern selection has been
thoroughly described in the detailed review [1] where one can find
numerous references.

The aim of the present paper is threefold: (i) To formulate a general
principle of pattern selection for nonequilibrium systems; (ii) To
derive its equivalent representations that could be convenient for
different cases; (iii) To emphasize the generality of the suggested 
principle by analysing different special applications.

\section{Pattern Distribution}

Consider a system of nonlinear evolution equations, which displays
the multiplicity of solutions corresponding to different patterns.
Let these solutions be parametrized by a multiparameter $\beta$
from a manifold ${\cal B}$.
For the simplicity of notation, let us examine, first, the case
when the considered problem can be reduced to a $d$-dimensional
system of ordinary differential equations written in the normal form
\begin{equation}
\label{1}
\frac{d}{dt}\; y(t) = v(y,t) \; ,
\end{equation}
where the state $y(t) =\{ y_i(t)=y_i(\beta,t)|\; i=1,2,\ldots,d\}$ and 
velocity field $v(y,t) =\{ v_i(y,t)| \; i=1,2,\ldots,d\}$ are the vectors,
in which $\beta\in{\cal B}$, and $d$ pertains to a countable set.
Evolution equations of many rather complicated systems can often
be presented in the form (1), including many continuous systems
after the reduction of their dynamics to the center manifold [2].
After examining the form (1), it will be straightforward to
generalize the consideration for the system of partial
differential equations.

As far as Nature does prefer some of the dynamical states,
labelled by $\beta$, as more probable, there should exist a
probability measure on the manifold ${\cal B}$. If we were able to
define a probability distribution $p(\beta,t)$ for a dynamical
system having at the moment $t$ a dynamical state
$y(t)=y(\beta,t)$, this would be actually the solution of the
problem of pattern selection: Then a pattern labelled by $\beta_1$
would be preferred over another one labelled by $\beta_2$ if
$p(\beta_1,t)> p(\beta_2,t)$. The most probable pattern would be
given by the {\it maximum of pattern distribution}.

According to the ideas of statistical mechanics [3], a probability
$p$ is related to entropy $S$ as $p\sim e^{-S}$. Or, since it is
not entropy itself but rather its variation $\Delta S$ that is
measurable, it is more appropriate to write $p\sim e^{-\Delta S}$.
Hence the probability distribution can be presented as
\begin{equation}
\label{2}
p(\beta,t) =\; \frac{1}{Z(t)}\; e^{-\Delta S(\beta,t)} \; , \qquad
Z(t) = \int_{\cal B} \; e^{-\Delta S(\beta,t)} \; d\beta \; .
\end{equation}
Thus, the most probable pattern corresponds to the {\it minimum of entropy 
variation}.

In order that the principle of pattern selection would be not just
a declaration but a working tool, it is necessary to express the
entropy variation through the {\it dynamical states} $y(\beta,t)$. For a
{\it nonequilibrium system}, the entropy variation can be naturally
defined as the difference
\begin{equation}
\label{3}
\Delta S(t) = S(t) - S(0)
\end{equation}
with respect to the initial time, which is a kind of relative
entropy [4]. Here and in the following intermediate expressions, we
shall omit, for brevity, the labelling multiparameter $\beta$,
always keeping in mind its existence and restoring it in final
formulas. The entropy may be defined as the logarithm of an
elementary phase volume, $S(t) =\ln |\delta\;\Gamma(t)|$;
the latter, for a {\it dynamical system}, being
$\delta\;\Gamma(t) =\prod_i \; \delta y_i(t)$.
Therefore, the entropy variation (3) writes
\begin{equation}
\label{4}
\Delta S(t) = \ln\left |\delta\;\Gamma(t)|{\Large /}\delta\;\Gamma(0)
\right | \; .
\end{equation}
Introducing [5] the multiplier matrix $\hat M(t)$, with the elements
$M_{ij}(t) \equiv\delta y_i(t)/\delta y_j(0)$, and initial condition 
$M_{ij}(0) = \delta_{ij}$, the elementary phase volume can be presented as
$\delta\Gamma(t) =\prod_i \; \sum_j  M_{ij}(t)\delta y_j(0)$,
from where $\Delta S(t) = \ln\left | \prod_i \sum_j \;
M_{ij}(t)\; M_{ji}(0)\right |$.
Taking into account the initial condition for the multiplier
matrix, the entropy variation (4) becomes
\begin{equation}
\label{5}
\Delta S(t) =\sum_i \; \ln|M_{ii}(t)| \; .
\end{equation}
Consequently, the pattern distribution (2) acquires the form
\begin{equation}
\label{6}
p(\beta,t) =\; \frac{1}{Z(t)} \; \exp\left\{ - \sum_i \;
\ln|M_{ii}(\beta,t)| \right\} = \; \frac{1}{Z(t)} \;
\prod_i \; |M_{ii}(\beta,t)|^{-1} \; .
\end{equation}
In this way, for an ensemble of dynamical states $y(\beta,t)$, one
may define the multiplier matrix and calculate the pattern
distribution (6).

In addition to expression (6), it is useful to give one
more representation for the pattern distribution. For this
purpose, let us introduce the matrix $\hat L(t)$ with the elements
$L_{ij}(t) \equiv \ln|M_{ij}(t)|$. Then the entropy variation (5) reads
\begin{equation}
\label{7}
\Delta S(t) = {\rm Tr}\;\hat L(t) \equiv \sum_i \; L_{ii}(t) \; .
\end{equation}
Varying the evolution equation (1), one gets the equation
\begin{equation}
\label{8}
\frac{d}{dt} \; \hat M(t) =\hat J(y,t)\; \hat M(t)\; , \qquad
J_{ij}(y,t) \equiv \; \frac{\delta v_i(y,t)}{\delta y_j(t)} \; ,
\end{equation}
for the multiplier matrix, where $\hat J$ is the Jacobian matrix.

To define the entropy variation (7), one, actually, does not need
to know the whole multiplier matrix but only the trace of the
matrix $\hat L(t)$. As far as the trace of a matrix does not depend on
its representation, one can accomplish intermediate
transformations in a chosen particular representation, returning
at the end to the form valid for arbitrary representations. Here,
at intermediate steps, we may consider the representation where
the multiplier matrix is diagonal. If so, equation (8) yields for the 
diagonal elements of the multiplier matrix
$M_{ii}(t) = \exp\left\{ \int_0^t J_{ii}(y(t'),t') dt' \right\}$.
Then, the diagonal elements of the matrix $\hat L(t)$ are
$L_{ii}(t) =\int_0^t \; {\rm Re} J_{ii}(y(t'),t') dt'$.
Let us introduce the quantity $K(t) \equiv \sum_i {\rm Re} J_{ii}(y,t)$.
Without the loss of generality, one may assume that the evolution
equation (1) is written for real functions, since an equation for
a complex function can always be presented as a pair of equations
for real functions. Hence the eigenvalues of the Jacobian matrix
are either real or, if complex, come in complex conjugate
pairs. This implies that
$\sum_i \; {\rm Re}\; J_{ii}(y,t) = {\rm Tr}\; \hat J(y,t)$.
Therefore the notation for $K$ can be written as
\begin{equation}
\label{9}
K(t) = {\rm Tr}\; \hat J(y,t) \; .
\end{equation}
The latter, in dynamical theory [6], is termed the {\it
contraction rate}. Using the equality
${\rm Tr}\; \hat L(t) = \int_0^t K(t') dt'$,
we find for the entropy variation (7)
\begin{equation}
\label{10}
\Delta S(t) = \int_0^t \; K(t')\; dt' \; .
\end{equation}

With Eq. (10), we get the pattern distribution
\begin{equation}
\label{11}
p(\beta,t) = \; \frac{1}{Z(t)} \; \exp\left\{
-\int_0^t \; K(t')\; dt'\right\} \; .
\end{equation}
Defining the {\it local contraction}
\begin{equation}
\label{12}
\Lambda(t) \equiv\; \frac{1}{t} \; \int_0^t \; K(t')\; dt' \; ,
\end{equation}
one may write $\Delta S(t) = \Lambda(t) t$.
Then the pattern distribution (11) takes the form
\begin{equation}
\label{13}
p(\beta,t) = \; \frac{1}{Z(t)} \; \exp\left\{ -
\Lambda(\beta,t)\; t\right\} \; .
\end{equation}
The latter shows that the most probable pattern is given by the
{\it minimum of local contraction}.

The generalization to the case when evolution equations (1) represent a set
of partial differential equations is straightforward. Then the dynamical
state $y(t)=\{ y_i(x,t)\}$ consists of functions of time as well as of a set
$x$ of continuous, say spatial, variables. Then $y(t)$ can be treated as a
vector with respect to the discrete index $i$ and to the continuous
multi-index $x$. The multiplier and Jacobian matrices are to be considered
as matrices in $i$ as well as in $x$, having the elements 
$$ 
M_{ij}(x,x',t)\equiv \; \frac{\delta y_i(x,t)}{\delta y_j(x',0)}\; , \qquad
J_{ij}(x,x',y,t) \equiv \; \frac{\delta v_i(x,y,t)}{\delta y_j(x',t)}\; ,
$$
the initial conditions for the multiplier matrix being
$M_{ij}(x,x',0) =\delta_{ij}\; \delta(x-x')$.
Thus, employing the matrix notation [5], we may literally
repeat the same steps as above, keeping in mind that, instead of one index
$i$, we have a pair of $i$ and $x$. Then the sums $\sum_i$ are to be
accompanied by the integrals $\int \; dx$. The product over a continuous
variable can be naturally defined [7] as 
$\prod_x f(x) \equiv\exp \int \ln f(x)dx$. 
As is clear, for the contraction rate (9), we have 
\begin{equation} 
\label{14} 
K(t) = \sum_i \; \int \;
J_{ii}(x,x,y,t) \; dx \; . 
\end{equation} 
The pattern distribution retains the same form (13), with the same local 
contraction (12), where the contraction rate is given by expression (14). 

In this way, we have shown that each dynamical state $y(\beta,t)$,
corresponding to a pattern labelled by a multi-index $\beta$, can
always be equipped with a weight defining the probability
distribution of patterns. The latter can be presented in several
equivalent forms as (2), (6), (11), or (13). Defining the
pattern weights makes it possible to organize a hierarchy among
different dynamical states of an ensemble $\{ y(\beta,t)|\; \beta\in
{\cal B}\}$ of admissible solutions. That state is more
preferable, which has a higher weight. The largest weight
describes the most probable pattern. It is also possible to define
the {\it average pattern} as ascribed to the average
$\overline\beta(t) \equiv \int_{\cal B}\beta p(\beta,t)d\beta$.
As for any statistical ensemble, one may define the {\it pattern
dispersion} $\sigma^2(t) \equiv \int_{\cal B}  \beta^2
p(\beta,t)  d\beta  - [\overline\beta(t)]^2$,
the standard deviation, variance coefficient, and so on.

The general principle for pattern selection is, briefly speaking,
the {\it maximum of pattern distribution}. This can be
reformulated in several forms according to a representation
employed. For instance, equation (13) shows that this principle
can be formulated as the {\it minimum of local contraction}. The
latter can be expressed either through the diagonal elements of
the multiplier matrix or through those of the Jacobiam matrix as
\begin{equation}
\label{15}
\Lambda(\beta,t) =\; \frac{1}{t} \; \sum_i \; \int \;
\ln|M_{ii}(x,x,\beta,t)|\; dx = \frac{1}{t} \; \int_0^t
{\rm Tr}\; \hat J(y(\beta,t'),\beta,t') \; dt' \; ,
\end{equation}
where trace implies summation over discrete indices and
integration over continuous variables. The conditions of minimum,
$\partial\Lambda(\beta,t)/\partial\beta = 0, \; 
\partial^2\Lambda(\beta,t)/\partial\beta^2 > 0$,
define $\beta(t)$ corresponding to a pattern preferable at time
$t$. {\it For nonequilibrium dynamical systems, the local
contraction plays the same ordering role as free energy for
equilibrium statistical systems}.

\section{Particular Cases}

In order to emphasize the generality of the advanced principle,
we shall consider several particular cases of nonequilibrium systems
having rather different properties.

\vskip 2mm

{\bf A. One-Dimensional Systems}. In this simplest case, the local 
contraction (12) is what is called the local Lyapunov exponent [9-13],
which can also be written as $\Lambda(t)=\frac{1}{t}\ln|M(t)|$.
The limit $\Lambda=\lim_{t\rightarrow\infty}\Lambda(t)$ is the
global Lyapunov exponent. Thus, the local contraction is closely 
connected with the stability properties of dynamical systems. And it 
becomes clear why a smaller local contraction defines a more probable
pattern. This is because a smaller local contraction corresponds to a more 
stable dynamical system.

\vskip 3mm

{\bf B. Hamiltonian Systems}. The dynamical state $y=\{ q,p\}$ consists of 
a pair of sets, $q=\{ q_i(x,t)\}$ and $p=\{ p_i(x,t)\}$, satisfying 
the system of  
Hamiltonian equations $\partial q/\partial t =\delta H/\delta p$ and
$\partial p/\partial t =-\delta H/\delta q$.
For the trace of the Jacobian matrix, one has
${\rm Tr}\hat J ={\rm Tr}\left (\delta^2 H/\delta p\delta q -
\delta^2 H/\delta q \delta p\right ) = 0$,
where the trace includes the summation over $i$ and integration over $x$, as
in Eq. (14). Consequently, $\Lambda=K=0$, which means that there are no 
multiple patterns but each pattern is to be uniquely defined by initial and 
boundary conditions.

\vskip 3mm

{\bf C. Chaotic Systems}.  In the spectrum of Lyapunov exponents, there are 
positive exponents. The sum of the latter is called the Kolmogorov-Sinai 
entropy [8]. The difference with the latter in our case is that the limit 
$\Lambda=\lim_{t\rightarrow\infty}\Lambda(t)$ is the sum of all Lyapunov 
exponents, but not of only positive ones. Hence, the principle of the 
local-contraction minimum makes it possible to classify even chaotic 
systems onto more or less stable.
It is worth emphasizing that the {\it local contraction}
characterizes the {\it local stability}. This is important since
the asymptotic, as $t\rightarrow\infty$, stability is known to be a
too rough notion for typical dynamical systems describing
realistic physical situations, as far as the phase spaces of these
systems are usually incredibly complicated, being composed of a
mixture of stability islands and chaotic domains [14,15].

\vskip 3mm

{\bf D. Dissipative Systems}. These, by definition, have negative 
contraction rates, $K={\rm Tr}\hat J<0$. The evolution equations for 
dynamical states $y=y(\beta,t)$ are often written in the form
$\partial y/\partial t =-\delta F/\delta y$, where 
${\rm Tr}\delta^2 F/\delta y^2 > 0$, with $F=F[y]$ being a functional of $y$.
Such a form of the evolution equation is typical, e.g., for the 
evolution of order parameters [16]. The local contraction (12) becomes
$$
\Lambda(\beta,t) =-\frac{1}{t} \int_0^t\frac{ {\rm Tr} 
\delta^2 F[y]}{\delta y^2(\beta,t')} \; dt' \; .
$$
In the process of evolution, the state $y$ tends to a solution minimizing 
the functional $F[y]$. If the latter possesses two or several minima, one 
encounters the so-called bistability or, respectively, multistability effects.
The minima of $F[y]$ are attractors of the dynamical system. Each attractor
is surrounded by its basin of attraction. For initial conditions inside a 
basin of attraction, the solution always tends to the corresponding attractor.
The problem of pattern selection arises when initial conditions are on the 
Julia set separating different basins of attraction. Then the solution may 
tend to different attractors. The probability of ending at the 
corresponding attractor, labelled by the index $\beta$, is given by the 
pattern distribution (13). The highest probability is defined by the 
minimum of the local contraction $\Lambda(\beta,t)$.

\vskip 3mm

{\bf E. Isolated Systems}. Such systems are characterized by nonnegative 
entropy production [17]. The latter, as follows from Eqs. (3) and (10), is 
directly expressed through the contraction rate
$K(\beta,t) =dS(\beta,t)/dt \geq 0$.
The second law of thermodynamics for an isolated system tells that the 
entropy does not decrease, that is $K\geq 0$. The system, as time 
increases, tends to an equilibrium state, so that $K(\beta,t)\rightarrow 
K(\beta)$ as $t\rightarrow\infty$. Then the local contraction (12) also 
tends to a constant, $\Lambda(\beta,t)\rightarrow K(\beta)$. The 
thermodynamic system at equilibrium acquires a structure providing the maximum
of entropy and the minimum of entropy production [17]. Labelling this 
structure by $\beta_0$, one has $K(\beta_0)=0$ and
$K'(\beta_0)=0$, with $K''(\beta_0)>0$, where primes imply the derivatives 
over $\beta_0$. Assume that in the process of evolution, a nonequilibrium 
system allows the existence of several structures classified by the pattern 
distribution (13). With time, one has
$$
p(\beta,t) \simeq \frac{1}{Z(\beta,t)}\; \exp\left\{ - K(\beta)\; t
\right \} \; , \qquad
Z(\beta,t)\simeq \int \exp\{ - K(\beta)\; t\} \; d\beta \; ,
$$
as $t\rightarrow\infty$. When the system tends to equilibrium, where the 
contraction rate is minimal, and $K(\beta_0)=0$, then, employing the 
Laplace method, we find $Z(\beta,t)\simeq\sqrt{2\pi}\gamma(t), \;
\gamma(t)\equiv[K''(\beta_0)t]^{-1/2}$.
Expanding $K(\beta)$ in the 
vicinity of its minimum, we have
$$
p(\beta,t)\simeq \; \frac{1}{\sqrt{2\pi}\;\gamma(t)} \; \exp\left\{ -\;
\frac{(\beta-\beta_0)^2}{2\gamma^2(t)}\right\}  \; .
$$
From here, we see that $p(\beta,t)\rightarrow\delta(\beta-\beta_0)$ as 
$t\rightarrow\infty$. This implies that an isolated system tending to 
equilibrium, even if it possessed the possibility of having several 
structures in the process of evolution, finally acquires the sole structure 
defined by the minimum of local contraction or by the maximum of entropy. 
The latter two conditions, for an equilibrium system, coincide.

\vskip 3mm

{\bf F. Filamentary Structures}. It would be instructive to mention an 
example for which theoretical results could be directly compared with 
experimental observations. A good case for this purpose is the problem of
the turbulent photon filamentation in resonant
media with high Fresnel numbers (see reviews [18,19] and
references therein). The problem of pattern formation in nonlinear
optics for high Fresnel numbers $F>10$ is principally different
from that for low Fresnel numbers $F<10$. In the latter case, all
arising patterns are uniquely described by the empty-cavity
Gauss-Laguerre modes. While at high Fresnel numbers the filament
patterns have nothing to do with these modes displaying a
nonunique variety of filaments with different radii. First, it has
been suggested [20-22] that the problem of pattern selection in
nonlinear optics at high Fresnel numbers can be treated being
based on the condition of minimal average energy, by analogy with
equilibrium systems. However, such a condition of minimal energy,
in general, has no grounds for nonequilibrium systems. 
The latter are to be treated by the approach suggested in this paper. For 
this purpose, let us consider a system of resonant two-level atoms. The 
variables describing interlevel transitions and the population difference 
are given by the statistical averages 
\begin{equation}
\label{16}
u({\bf r},t) \equiv 2<S^-({\bf r},t)>\; , \qquad
s({\bf r},t) \equiv 2<S^z({\bf r},t)>
\end{equation}
of the quasispin operators [23]. The evolution equations for these 
quantities can be derived by invoking the method of eliminating field 
variables [19,23], which yields
$$
\frac{\partial u}{\partial t} = - (i\omega_0 +\gamma_2) u + fs \; ,
\qquad
\frac{\partial |u|^2}{\partial t} = -2\gamma_2|u|^2 +(u^*f + f^*u) s \; ,
$$
\begin{equation}
\label{17}
\frac{\partial s}{\partial t}=-\; \frac{1}{2} (u^*f + f^* u) -
\gamma_1 (s -\zeta ) \; .
\end{equation}
Here $\omega_0$ is the atomic transition frequency, $\gamma_1$ and 
$\gamma_2$ are the longitudinal and transverse  attenuation parameters, 
$\zeta>0$ is a pumping parameter, and
$$
f({\bf r},t) = -2i{\bf d}\cdot{\bf E}_0({\bf r},t) + f_{rad}({\bf r},t) \; ,
\qquad
\varphi({\bf r}) \equiv \; \frac{\exp(ik_0|{\bf r}|)}{k_0|{\bf r}|} \; ,
$$
$$
f_{rad}({\bf r},t) = -\; \frac{3}{4}\; i\gamma\rho\;
\int\left [ \varphi({\bf r} -{\bf r}')\; u({\bf r}',t) -
{\bf e}_d^2\varphi^*({\bf r}-{\bf r}') \; u^*({\bf r}',t)\right ]\;
d{\bf r}' \; ,
$$
with $\gamma\equiv (4/3)k_0^3 d_0^2, \;  k_0\equiv\omega_0/c$,
$\rho$ being the density of atoms, and ${\bf d}\equiv d_0{\bf e}_d$ being 
the transition dipole. The seed field
${\bf E}_0 =\frac{1}{2}\; {\bf E}_1\; e^{i(kz-\omega t)} +
\frac{1}{2}{\bf E}_1^*\; e^{-i(kz-\omega t)}$
selects the longitudinal mode with $\omega=kc$ and with a small detuning 
from the resonance, $|\Delta|\ll\omega_0, \; \Delta\equiv\omega-\omega_0$.
The evolution equations (17) are nonlinear integro-differential equations 
which may possess several solutions. Despite that the pumping is uniform, 
there may appear self-organized transverse modes visible as radiating 
filaments. The solutions describing a filamentary structure can be 
presented as the sums over $N_f$ filaments,
\begin{equation}
\label{18}
u({\bf r},t) =\sum_{n=1}^{N_f} u_n(r_\bot,t)e^{ikz} \; , \qquad
s({\bf r},t) =\sum_{n=1}^{N_f} s_n(r_\bot,t) \; ,
\end{equation}
where $r_\bot\equiv\sqrt{x^2+y^2}$ and the functions $u_n$ and $s_n$ are 
assumed to be essentially nonzero around the axis of an $n$-th filament but 
fastly decreasing outside the filament, so that $u_mu_n\sim \delta_{mn}$,
$s_ms_n\sim\delta_{mn}$, and $u_ms_n\sim \delta_{mn}$. Substituting 
expansions (18) into Eqs. (17), we obtain the evolution equations for the 
filament functions $u_n$ and $s_n$. These equations compose an 
infinite-dimensional dynamical system. To simplify the problem, we may pass 
from the infinite-dimensional system to its center manifold that would be 
of finite dinemsionality. To this end, let us introduce the averaged functions
\begin{equation}
\label{19}
u(t) \equiv \frac{1}{V_n}\; \int_{V_n} u_n(r_\bot,t)\; d{\bf r} \; , \qquad
s(t) \equiv \frac{1}{V_n}\; \int_{V_n} s_n(r_\bot,t)\; d{\bf r} \; , 
\end{equation}
with the averaging accomplished over the cylinder enveloping the $n$-th 
filament. The volume of the enveloping cylinder is $V_n =\pi b_n^2 L$, 
where $b_n$ is the cylinder radius and $L$ is the length of the sample. The 
enveloping cylinder radius $b_n$ is related to the filament radius $r_n$ by 
the conservation-energy relation
$\int |u_n(r_\bot,t)|^2d{\bf r} = V_n |u_n(r_n,t)|^2$.
This relation envolves the function $|u_n|^2$ since a filament as such is 
defined by its radiation intensity which is proportional to $|u_n|^2$. 
Assuming that the profile of the function $|u_n(r_\bot,t)|^2$ is well 
approximated by the normal law $\exp(-r^2_\bot/2r_n^2)$, we obtain
\begin{equation}
\label{20}
b_n=(4e)^{1/4}\; r_n = 1.82\; r_n \; .
\end{equation}
The whole sample is supposed to have the cylindrical shape of radius $R$ 
and length $L$, with the relation $R\ll L$ typical of lasers. The radiation 
wavelength is $\lambda\ll R$. The evolution equations for the averages 
(19) take the form
$$
\frac{d u}{d t} = - (i\Omega +\Gamma) u + f_1s \; , \qquad
\frac{d |u|^2}{d t} = -2\Gamma |u|^2 +(u^*f_1 + f^*_1 u ) s \; ,
$$
\begin{equation}
\label{21}
\frac{d s}{d t}= -g\gamma_2|u|^2 - \; \frac{1}{2} (u^*f_1 + f^*_1 u) -
\gamma_1 (s -\zeta ) \; ,
\end{equation}
where $\Omega\equiv \omega_0 +\gamma_2g' s$ is the collective frequency, 
$\Gamma\equiv\gamma_2(1-gs)$ is the collective width, $f_1\equiv -i 
{\bf d}\cdot{\bf E}_1 e^{-i\omega t}$, and the effective coupling 
parameters are \begin{equation}
\label{22}
g\equiv \frac{3\gamma\rho}{4\gamma_2}\; 
\int_{V_n} \frac{\sin(k_0r-kz)}{k_0r}\; d{\bf r}\; ,
\qquad
g'\equiv \frac{3\gamma\rho}{4\gamma_2}\; 
\int_{V_n} \frac{\cos(k_0r-kz)}{k_0r}\; d{\bf r}\; .
\end{equation}
Equations (21) can be solved by invoking the scale separation approach 
[24]. Taking into account the standard inequalities $\gamma\ll\omega_0$,
$\gamma_1\ll\omega_0$, and $\gamma_2\ll\omega_0$, the solutions to Eqs. (21)
can be classified onto fast, $u$, and slow, $|u|^2$ and $s$. Treating the 
slow functions as quasi-invariants, for the fast solutions one has
$$
u(t) =\left ( u_0 - \; \frac{s{\bf d}\cdot{\bf E}_1}{\omega-\Omega +i\Gamma}
\right )\; e^{-(i\Omega +\Gamma)t} + 
\frac{s{\bf d}\cdot{\bf E}_1}{\omega-\Omega+i\Gamma}\; e^{-i\omega t} \; .
$$
Substituting this into the equations for the slow functions, we average the 
right-hand sides of these equations over time and introduce the function
\begin{equation}
\label{23}
w \equiv |u|^2 - \alpha s^2 \; ,
\qquad
\alpha\equiv \frac{{\rm Re}}{s\gamma} \; \lim_{\tau\rightarrow\infty} \;
\frac{1}{\tau} \; \int_0^\tau u^*(t) f_1(t)\; dt =
\frac{|{\bf d}\cdot{\bf E}_1|^2}{(\omega-\Omega)^2+\Gamma^2} \; .
\end{equation}
The role of the seed field is to select a longitudinal mode, the amplitude 
$E_1$ being small, so that $\alpha\ll 1$. Finally, for the slow functions, 
we obtain the equations
\begin{equation}
\label{24}
\frac{d w}{dt} = -2\gamma_2(1-gs)w \; , \qquad
\frac{d s}{dt} = -2\gamma_2w -\gamma_1(s -\zeta) \; ,
\end{equation}
defining the guiding-center solutions. From here, we have the contraction 
rate, given by Eq. (9), as $K=-\gamma_1-2\gamma_2(1-gs)$. The latter 
depends on the filament radius through relations (20) and (22). Therefore, 
the pattern distribution (11) gives the distribution of photon filaments 
with respect to their radii. The most probable filament radius corresponds 
to the maximum of the pattern distribution (11), that is, to the minimum 
of the local contraction (12). Filaments with different radii have 
different values of coupling parameters (22). The stability analysis of 
Eqs. (24) shows that all filaments, independently of their radii, are 
stable from the point of view of the asymptotic Lyapunov stability. The 
classification of filaments onto more or less probable happens at the 
initial stage of their formation, when the local contraction (12) is
\begin{equation}
\label{25}
\Lambda(t) \simeq -\gamma_1 -2 \gamma_2 (1 -gs_0) \; ,
\end{equation}
where $s_0\equiv s(0)<0$. The minimum of $\Lambda$ corresponds to the 
maximum of $g$. Minimizing Eq. (25) with respect to the filament radius, we 
find the most probable radius $r_f=0.3\sqrt{\lambda L}$, which is in good 
agreement with experiments (reviewed in Ref. [19]) for various laser media.

\vskip 3mm

{\bf G. Time Series}. The general principle described in Sec. 2 can also be 
employed for analyzing time series by providing a probabilistic 
distribution of extrapolated scenarios. This is possible since a time 
series is a realization of a random process corresponding to a stochastic 
dynamical system. Different realizations of a random process can be 
presented by different time series (which we may enumerate by the index 
$\beta=1,2,\ldots$) representing the same random process. This implies that 
there are several sets of data $f_k(\beta)$ associated with the moments of 
time $t_k(\beta)$, with $k=0,1,2,\ldots$. Each set forms the data base
$\Bbb{D}_k(\beta) =\{ f_k(\beta),f_{k-1}(\beta),\ldots,f_0|\;
t_k(\beta)<t_{k-1}(\beta)<\ldots < 0\}$,
where the backward time ordering is accepted and the present time moments 
is set to be $t_0(\beta)=0$. To extrapolate a time series means to 
construct a forecast, valid for $t>0$, on the base $\Bbb{D}_k(\beta)$ 
containing the data for $t\leq 0$. Assuming that the considered random process 
follows the self-similar dynamics [25], we obtain the self-similar forecast
$$
f_k^*(\beta,t) = f_0\exp\left ( c_1 t\exp\left ( c_2 t\ldots\exp\left ( c_k t
\right )\right ) \ldots \right ) \; ,
$$
extrapolating the time series to $t>0$. Here $c_k=c_k(\beta,t)$ are control 
functions, or controllers, defined by minimizing a cost functional. Each 
forecast $f_k^*(\beta,t)$ presents a possible scenario for the extrapolated 
behaviour of the considered time series. In the present case, scenario is a 
synonym for pattern. Hence, we may construct a scenario probability
$p_k(\beta,t) = 1/Z_k(t)\left | M_k(\beta,t)\right |$
by using definition (6). Treating the family $\{ f_k^*(\beta,t)|\; 
k=1,2,\ldots\}$ as the trajectory of a dynamical system with discrete time 
$k=1,2,\ldots$, we find the multipliers
$M_k(\beta,t) \equiv\delta f_k^*(\beta,t)/\delta f_1^*(\beta,t)$.
Introducing the average multiplier $\overline
M_k(t)$ by the relation 
$Z_k(t) \equiv 1/\overline M_k(t) = \sum_\beta 1/|M_k(\beta,t)|$,
we obtain the scenario distribution
$$
p_k(\beta,t) = \frac{\overline M_k(t)}{|M_k(\beta,t)|} =
\overline M_k(t)\exp\left\{ - \Lambda_k (\beta,t)\; t\right \}
$$
with the local contraction $\Lambda_k=\frac{1}{t}\ln|M_k|$, in agreement 
with Eq. (15). Thus, the most probable scenario is defined by the minimum 
of the local contraction $\Lambda_k(\beta,t)$ or, respectively, by the 
minimum of the multiplier modulus $|M_k(\beta,t)|$. Several examples of 
particular time series are considered in Ref. [25], where a variant of the 
scenario distribution was postulated.

\vskip 3mm

In conclusion, a general principle for pattern selection is advanced, based 
on the definition of the probability distribution of patterns. It is 
demonstrated that the most probable pattern corresponds to the {\it minimum 
of local contraction}. The suggested approach is shown to be applicable to 
different dynamical systems describing various nonequilibrium phenomena.

\end{document}